\documentclass{article}

\usepackage[english]{babel}

\usepackage[letterpaper,top=2cm,bottom=2cm,left=3cm,right=3cm,marginparwidth=1.75cm]{geometry}

\usepackage{amsmath}
\usepackage{graphicx}
\usepackage[colorlinks=true, allcolors=blue]{hyperref}
\usepackage{subcaption}

\title{Comprehensive study of good model training for prostate segmentation in volumetric MRI}
\author{Carlos Nácher Collado}

\begin{document}
\maketitle

\begin{abstract}
Prostate cancer was the third most common cancer in 2020 internationally, coming after breast cancer and lung cancer. Furthermore, in recent years prostate cancer has shown an increasing trend. According to clinical experience, if this problem is detected and treated early, there can be a high chance of survival for the patient. One task that helps diagnose prostate cancer is prostate segmentation from magnetic resonance imaging. Manual segmentation performed by clinical experts has its drawbacks such as: the high time and concentration required from observers; and inter- and intra-observer variability. This is why in recent years automatic approaches to segment a prostate based on convolutional neural networks have emerged. Many of them have novel proposed architectures. In this paper I make an exhaustive study of several deep learning models by adjusting them to the task of prostate prediction. I do not use novel architectures, but focus my work more on how to train the networks. My approach is based on a ResNext101 3D encoder and a Unet3D decoder. I provide a study of the importance of resolutions in resampling data, something that no one else has done before. 

\end{abstract}

\section{Introduction}

According to the following study by the International Agency for Research on Cancer \cite{Sung2021GlobalCS}, in 2020 there were, globally, 1 414 259 new cases of prostate cancer, being the third most common cancer (7.3\%) after breast cancer (11.7\%) and lung cancer (11.3\%); referring these percentages to the total number of new cases of any cancer in 2020 internationally, about 19.3 million; producing 375 304 (3.8\%) deaths in that year. Clinical experience shows that if prostate cancer is detected early there is a high chance of survival for the patient. If prostate cancer is identified early, there is a high probability of survival for the patient. For this and other applications, automatic MRI prostate segmentation is useful for clinical support. In this paper I propose a comprehensive study of the training of several Deep Learning models. 

Automatic prostate segmentation in magnetic resonance imaging (MRI) with Deep Learning has been based since its inception on Convolutional Neural Networks (CNN) first proposed by \cite{Fukushima1975-xm} and later improved upon by
\cite{Lecun1998}. The best known CNN for medical image segmentation is the U-Net \cite{unet}. Since the \href{https://promise12.grand-challenge.org/}{PROMISE12} competition started in 2012, there have been several proposals for improvements: \textit{Haozhe Jia et al.} \cite{HJIA20181} proposed to do Transfer Learning \cite{transferlearning1976} of the VGG-19 \cite{vgg} network by separating the data according to whether the patient has endorectal coil or not; their coarse-to-fine strategy is based on atlas-based segmentation \cite{atlasbasedseg}. In 2018 \textit{Haozhe Jia et al.} \cite{HJIA20182} propose an improvement their previous contribution; they change the VGG-19 backbone \cite{vgg} to ResNet-50 \cite{resnet} in 3D; they propose the use of GCB blocks \cite{gcn} and RRB; furthermore, they propose the use of a Generative Adversarial Network (GAN) \cite{goodfellow2014generative} to discriminate the classification of each voxel and refine the results of the \textit{encoder-decoder}. In January 2019, \textit{Isensee et al.} \cite{nnunet} present a paper in which they claim that a well-trained basic U-Net network can achieve great results. They start from the 3D U-Net \cite{unet} \cite{3dunet} and say that instead of focusing on the architecture of the network and adding complex layers such as a gan, they focus on the correct training of the network; this premise of Isensee et al. is the inspiration for my work. In August 2019, \textit{Zhu et al.} \cite{bowda} propose an adaptation of the data domain with a GAN-like discriminator. In July 2019, \textit{Haozhe Jia et al.} \cite{HJIA3dapa} modify their GCBs and RRBs by Py-Conv and As-Conv, to account for the anisotropy of MRI data as well as extract features at various levels of resolution; they also introduce CABs (attention modules) and a specialised prostate edge decoder. In 2019, \textit{Xiangxiang Qin} \textit{HJIA3dapa} proposes a structure called PAM that combines the PyConv and the CABs of \textit{HJIA3dapa} into a single module; it also proposes a decoder specialising in the edges, but in 3D. In June 2022, and \textbf{being the state of the art as of today}, \textit{Haozhe Jia et al.} \cite{JIA2022} again propose an improvement to their previous contributions \cite{HJIA20181}, \cite{HJIA20182}, \cite{HJIA3dapa}; they change the ResNet backbone \cite{resnet} to Res2Net \cite{res2net} 

State of the art proposals, in due course, have introduced new concepts such as: the use of GANs \cite{goodfellow2014generative} \cite{HJIA20182} \cite{HJIA3dapa}; the use of convolutions at different scales \cite{HJIA3dapa} \cite{hdnet} \cite{alexnet} \cite{JIA2022}; the use of anisotropic convolutions \cite{HJIA20182} \cite{HJIA3dapa} \cite{hdnet} \cite{alexnet} \cite{JIA2022}; the use of attention modules \cite{hdnet} \cite{alexnet} \cite{JIA2022}; the use of a specialised decoder at prostate edges \cite{hdnet} \cite{JIA2022}.

However, there is one particular paper, \textit{No-New Net, by Isensee et al.} \cite{nnunet}(more extended in \textit{Isensee et al.} \cite{Isensee2019nnUNetBT}), which hypothesises that a "basic" network for 3D segmentation, such as the 3D-Unet \cite{3dunet}, may be sufficient to achieve state-of-the-art results. He focuses his research on how to train the neural network, and not on the architecture itself. He has proven to be among the top performers on several occasions with his work \textit{Isensee et al.} \cite{nnunet} and is currently ranked fourth on \href{https://promise12.grand-challenge.org/}{PROMISE12}.

Based on this work, \textbf{I propose an intensive study of the training} of a basic network (3D-Unet \cite{3dunet}), with different tests on: the choice of the backbone used; the optimiser used and its hyperparameters; the Data Augmentation techniques used. In addition, I propose a study on the way of resampling the data and the choice of the common output resolution, \textbf{which no one else has proposed in their work, with which I show that there is room for improvement in other aspects of the task of training an Artificial Intelligence model, other than the choice of the architecture itself}.

\section{Method}

\subsection{Implementation}

I use the 50 training cases provided by the contest. As preprocessing I do MCLAHE \cite{mclahe}.
As data augmentation I do flips in the x,y,z axis; rotations from -15 to 15 degrees in the z axis; scaling between 0.75 and 1.5; and BSpline deformations. All these transformations are applied online to random patches of 192x192x32. 

The loss function used was Dice \ref{dice} + BinaryFocalCrossentropy \ref{bfce}. The optimizer used was Adam with initial learning rate of 0.006, with decay when validation loss stopped improving and epsilon = 0.001.

All training was run on a 16GB Tesla P100 GPU graphics card.

\subsection{Comprehensive study of training}

The first thing I propose is a metric called Reconstruction Dice (rDSC). Basically this means, resampling a segmentation to a particular resolution different from its original resolution, resampling it, and then calculating the Dice coefficient between the original segmentation and the reconstructed segmentation. I do this to study the resampling of the training data to the optimal resolution. The resolutions I propose for the 50 training cases are:

\begin{itemize}
    \item Half of the resolution of each particular segmentation.
    \item Twice the resolution.
    \item The median of the resolutions in the training set (0.625x0.625x3.6).
    \item The resolution (0.625x0.625x1.5) as this is the resolution used by most of the PROMISE12 papers.
    \item A proposal (0.3125x0.3125x1.5).
\end{itemize}

The proposed resolution (0.3125x0.3125x1.5) was proposed because it was observed that decreasing the resolution of a piece of data (i.e. making its spacing larger) resulted in a worse rDSC than increasing the resolution (i.e. making its spacing smaller). This is because reducing the resolution often discards information that, when you want to return to the original resolution, is not available because it was lost in the first resampling. On the other hand, when the resolution is increased, new values are "invented" (interpolated), and then it is easier to return to the original image with fewer losses.

\begin{figure}[!htb]
    \centering
    \includegraphics[width=0.95\textwidth]{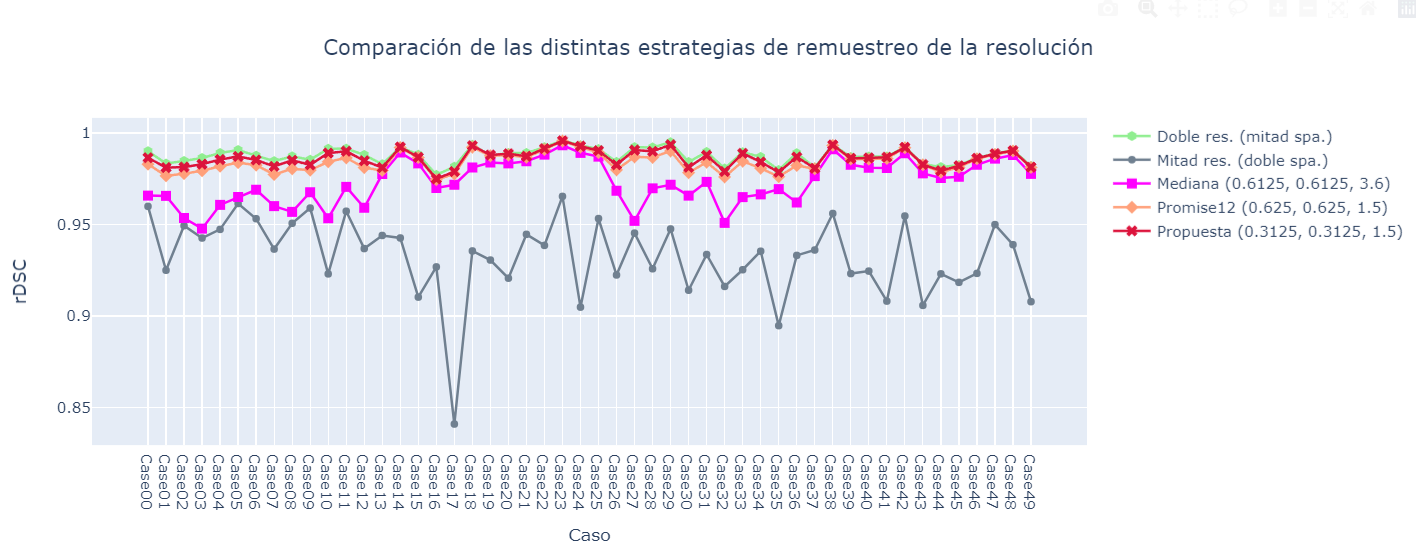}
    \caption{Comparison of the different resolutions in terms of rDSC. Note that resampling each image to twice its resolution is ideal in terms of rDSC but with 512x512xz images, this strategy would leave us with 1024x1024x2z data, which is computationally unfeasible. The proposed strategy outperforms the standard strategy used in PROMISE (0.625x0.625x1.5) with p-value $<$ 0.05.
    }
    \label{fig:resolutions_train} 
\end{figure}

I tested all five strategies with the 50 training cases. The results can be seen in figure \ref{fig:resolutions_train}. With a p-value $<$ 0.05, the proposed (0.3125x0.3125x1.5) achieves better rDSC than (0.625x0.625x1.5). Later on, I try to train the best of my models for data with both resolutions, and show that the one using data with resolutions (0.3125x0.3125x1.5) obtains better validation metrics. (See \ref{fig:ablation})

Then, with the resolution (0.3125x0.3125x1.5) chosen, I make some intensive tests to choose the data augmentation to use, the cost function, the backbone and the decoder model, among others.

In terms of cost function, I tried 5: 

\begin{itemize}
    \item Dice \ref{dice}
    \item BinaryFocalCrossEntropy \ref{bfce}
    \item Tversky \ref{tversky}
    \item \textbf{Dice+ BinaryFocalCrossEntropy} 
    \item Tversky + BinaryFocalCrossEntropy
\end{itemize}

The best one turns out to be \textbf{Dice+ BinaryFocalCrossEntropy}, so it is the one I choose to train the models. 

\begin{equation} \label{dice}
\text { Dice }=\frac{2|A \cap B|}{|A|+|B|}
\end{equation}

\begin{equation} \label{bfce}
CE = -\frac{1}{n} \sum_{j=1}^{n} \sum_{i=1}^{c}\left[y_{i} \log \left(p_{i}\right)+\left(1-y_{i}\right) \log \left(1-p_{i}\right)\right]
\end{equation}

\begin{equation} \label{tversky}
\mathrm{TI}=\frac{\mathrm{TP}}{\mathrm{TP}+\alpha \mathrm{FN}+\beta \mathrm{FP}}
\end{equation}

As backbones try several (freezing weights vs. non-freezing):

\begin{itemize}
    \item ResNet18 \cite{resnet}
    \item ResNet50 \cite{resnet}
    \item \textbf{ResNext101} \cite{res2net}
\end{itemize}

\begin{figure}[!htb]
    \centering
    \includegraphics[width=0.7\textwidth]{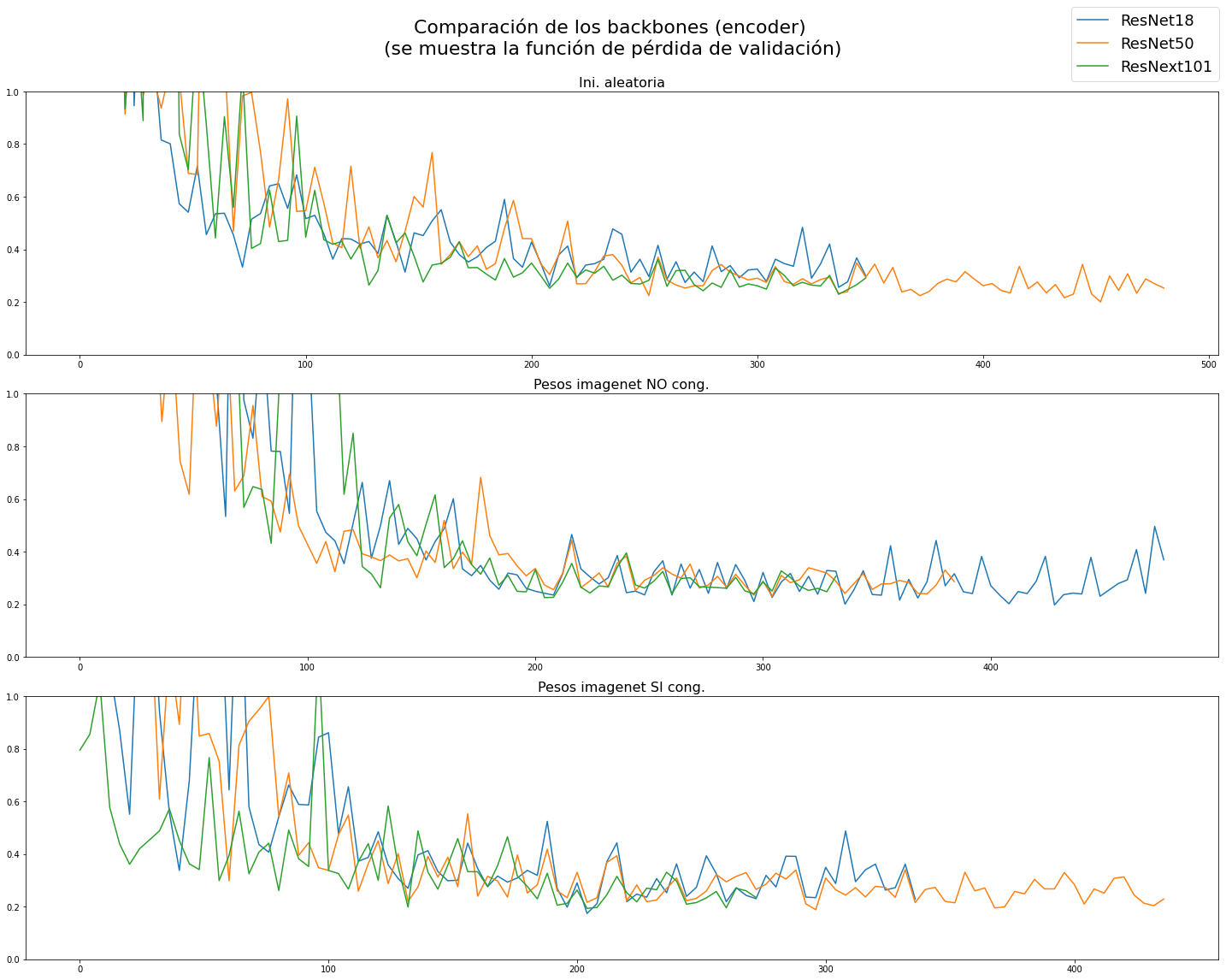}
    \caption{Comparison of backbones. The losses are a bit noisy because they are the first epochs.
    }
    \label{fig:backbones} 
\end{figure}

\begin{figure}[!htb]
    \centering
    \includegraphics[width=0.5\textwidth]{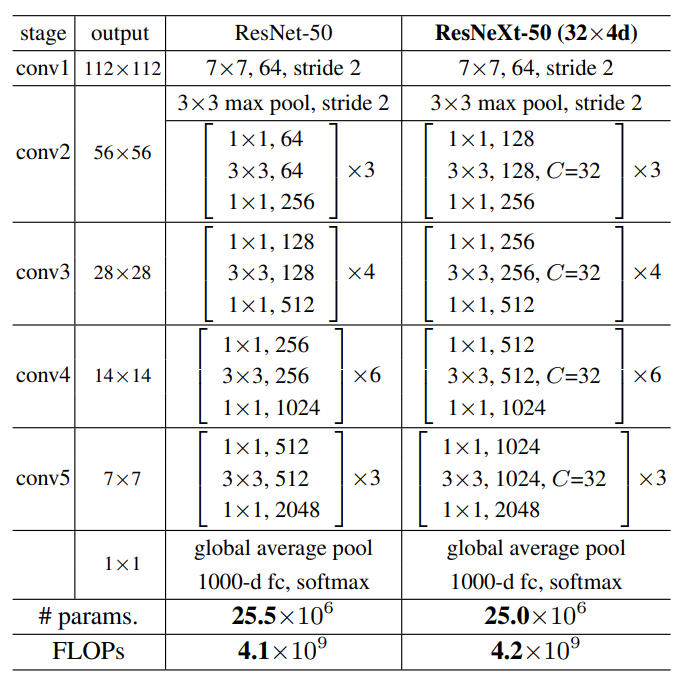}
    \caption{ResNet and ResNext architetures and shapes.
    }
    \label{fig:resnet_archs} 
\end{figure}

\begin{figure}[!htb]
    \centering
    \includegraphics[width=0.7\textwidth]{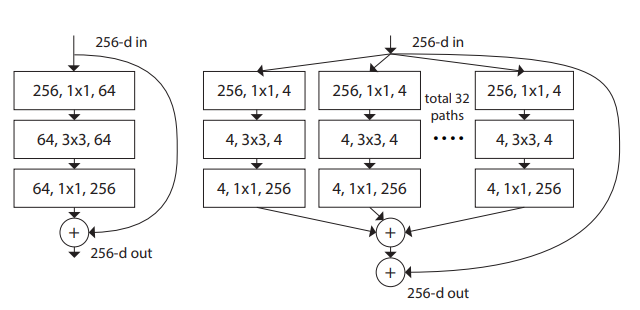}
    \caption{Comparison of ResNet vs ResNext block.
    }
    \label{fig:resnet_blocks} 
\end{figure}

A comparison of the tested backbones can be seen in figure \ref{fig:backbones}. The results are somewhat noisy, but according to the T-test, ResNext101 is better with p-value $<$ 0.05. The general architecture of ResNext can be seen in figure \ref{fig:resnet_archs}, and a comparison between ResNet and ResNext blocks (the latter being equivalent but more efficient) in figure \ref{fig:resnet_blocks}.

\begin{figure}[!htb]
     \centering
     \begin{subfigure}[b]{0.3\textwidth}
         \centering
         \includegraphics[width=1.5\textwidth]{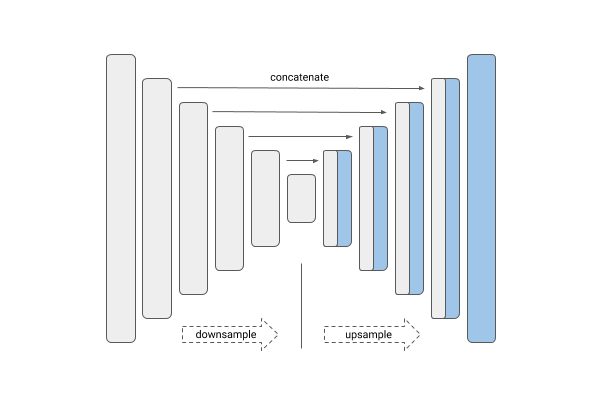}
         \caption{U-Net architecture.}
         \label{fig:unet}
     \end{subfigure}
     \hfill
     \begin{subfigure}[b]{0.3\textwidth}
         \centering
         \includegraphics[width=1.5\textwidth]{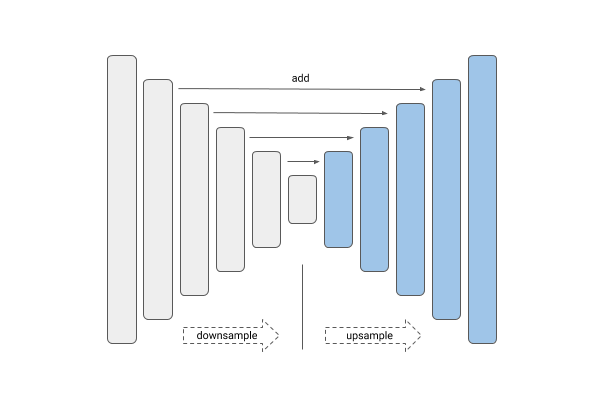}
         \caption{Link-Net architecture.}
         \label{fig:linknet}
     \end{subfigure}
     \hfill
     \begin{subfigure}[b]{0.3\textwidth}
         \centering
         \includegraphics[width=1.5\textwidth]{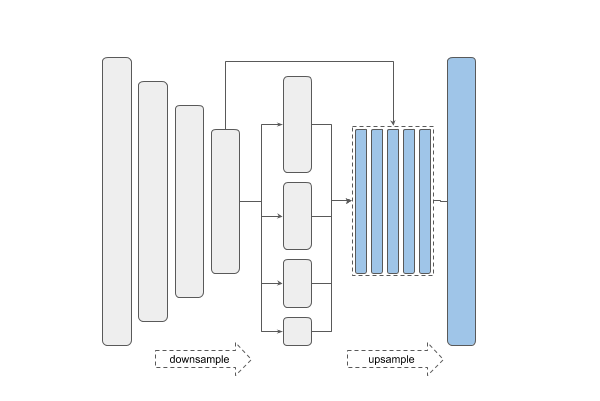}
         \caption{PSP-Net architecture.}
         \label{fig:pspnet}
     \end{subfigure}
     \caption{Comparison of decoder model architetures.}
     \label{fig:models} 
\end{figure}

As decoder models I test: the U-Net 3D \cite{unet} \cite{3dunet}, the LinkNet 3D \cite{linknet} (which is identical to the U-Net with the difference of adding instead of concatenating) and the PSP-Net \cite{pspnet}. The schematic of the three architectures is shown in figure \ref{fig:models}. The U-Net 3D gave me the best results.

\begin{figure}[!htb]
    \centering
    \includegraphics[width=0.7\textwidth]{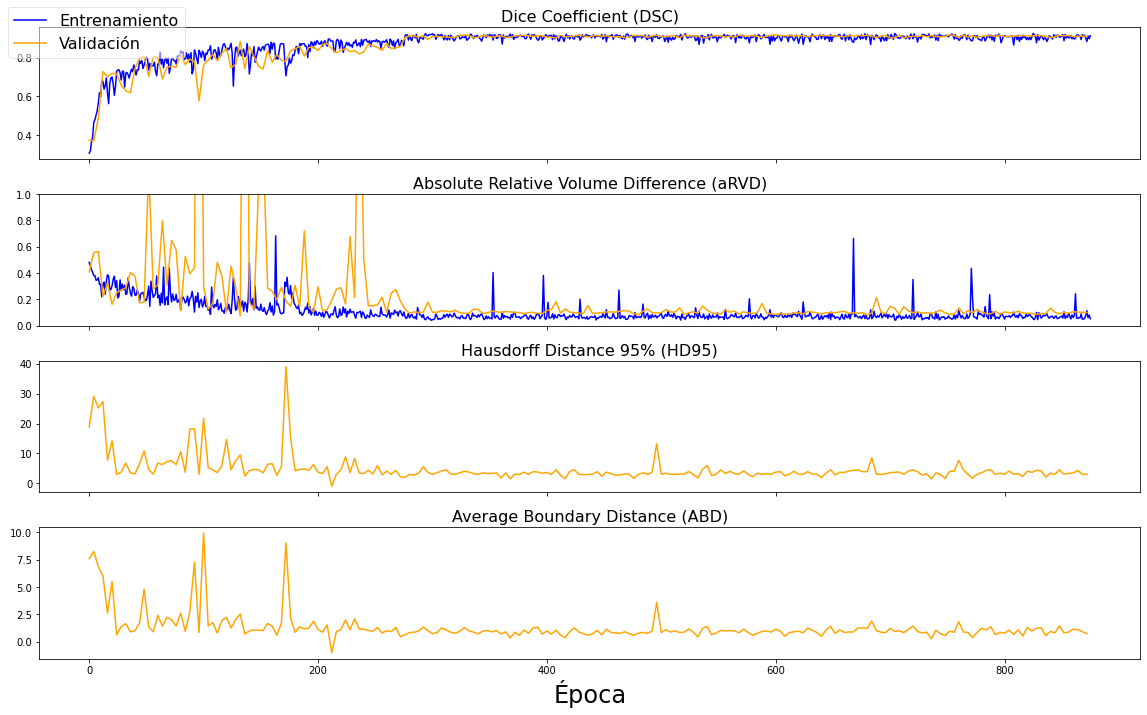}
    \caption{Metrics on train and validation with the best model achieved.
    }
    \label{fig:diceval} 
\end{figure}

Tras esta y otras muchas pruebas que no incluyo en este paper, consigo el mejor modelo, cuyo Dice de validación muestro en la figura \ref{fig:diceval}.

For predictions I use the Sliding Windows technique with windows of size 192x192x32 and fixed stride of 48x48x8, which gives me better results than using stride of 96x96x16.

\section{Ablation experiments}

\begin{figure}[!htb]
    \centering
    \includegraphics[width=0.7\textwidth]{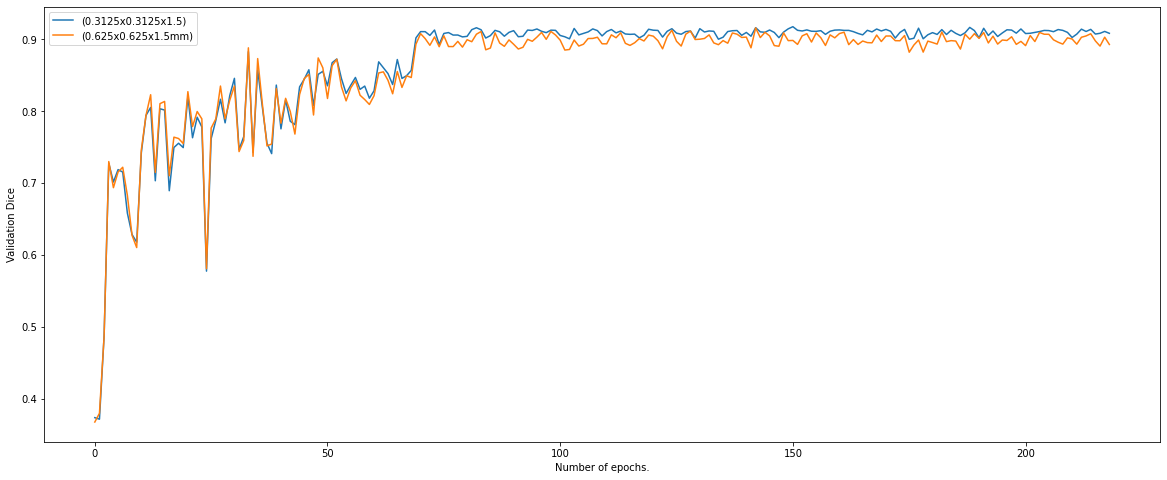}
    \caption{Comparison of training the same models with data in resolution (0.625x0.625x1.5mm) vs (0.3125x0.3125x1.5mm).
    }
    \label{fig:ablation} 
\end{figure}

Once I got the best model for the resolution (0.3125x0.3125x1.5), I resampled the data again, this time to 0.625x0.625x1.5mm, and trained the best model but now with the new data (keeping all the hyperparameters like the batch size the same, also to occupy the same physical space I took the random patches of size 96x96x32). In figure \ref{fig:ablation} you can see how the proposed strategy improves the validation Dice by up to one point. 

\section{Conclusions}

I have shown that good training of a basic network can contribute sufficiently good results compared to other complex methods, just as Isensee et al. demonstrate with \cite{nnunet}.

I have shown that there is room for improvement of the algorithms if we focus on the way the data is handled and not only on the architecture. Demonstrating how the resampling resolution of the data can improve Die by up to 1 point.

I am waiting for the test results at the hands of the \href{https://promise12.grand-challenge.org/}{PROMISE12} competition.

\bibliographystyle{alpha}
\bibliography{main}

\end{document}